\documentclass [6pt,a4paper]{article}
\usepackage{bbm}
\usepackage{amsfonts}
\usepackage{mathrsfs}
\usepackage {amssymb}
\usepackage {amsmath}
\usepackage{amsthm}
\usepackage{latexsym}
\usepackage{graphicx}
\usepackage[english]{babel}
\usepackage[latin1]{inputenc}
\usepackage{times}
\usepackage[T1]{fontenc}
\usepackage{CJK}
\usepackage{amscd}

\def\dse#1{\vskip 0.6cm\noindent
        {\large\bf #1}
        \vskip 0.4cm}
\newcommand{\pf}{{\bf Proof. \ }}
\def\dse#1{\vskip 0.6cm\noindent
        {\large\bf #1}
        \vskip 0.4cm}
\usepackage{lastpage}
\usepackage{epsfig}

\begin{document}
\begin{center}
\textbf{\large{ A family of constacyclic codes over
$\mathbb{F}_{2^{m}}+u\mathbb{F}_{2^{m}}$ and its application to
quantum codes}} \footnote { This research is supported by the
Natural Science Foundation of Anhui Province (No. 1808085MA15), Key
University Science Research Project of  Anhui Province (No.
KJ2018A0497), National Natural Science Foundation of China (Nos.
61772168 and
61572168).\\
~~$^\star$Corresponding author.\\
~~E-mail addresses: ysh$_{-}$tang@163.com(Y.Tang),yaoting$_{-}$1649@163.com(T.Yao), sxinzhu@tom.com(S.Zhu),kxs6@sina.com(X.Kai).
}

\end{center}

\begin{center}
{ { Yongsheng Tang$^{1\star}$, \ Ting Yao$^{2}$, \ Shixin Zhu$^{2}$,
\ Xiaoshan Kai$^{2}$} \ }
\end{center}

\begin{center}
\textit{\footnotesize $^{1}$Department of Mathematics, Hefei Normal
University, Hefei 230601,  Anhui,  P.R.China  \\
 $^{2}$ School of
Mathematics, Hefei University of Technology, Hefei 230601, Anhui,
P.R.China }\\
\end{center}

\noindent\textbf{Abstract:} We introduce a Gray map from
$\mathbb{F}_{2^{m}}+u\mathbb{F}_{2^{m}}$ to $\mathbb{F}_{2}^{2m}$
and study $(1+u)$-constacyclic codes over
$\mathbb{F}_{2^{m}}+u\mathbb{F}_{2^{m}},$ where $u^{2}=0.$ It is
proved that the image of a $(1+u)$-constacyclic code  length $n$
over $\mathbb{F}_{2^{m}}+u\mathbb{F}_{2^{m}}$ under the Gray map is
a distance-invariant quasi-cyclic code of  index $m$ and length
$2mn$ over $\mathbb{F}_{2}.$ We also prove that every code of length
$2mn$ which is the Gray image of  cyclic codes over
$\mathbb{F}_{2^{m}}+u\mathbb{F}_{2^{m}}$ of length $n$ is
permutation equivalent to a binary quasi-cyclic code of index $m.$
Furthermore,  a family of quantum error-correcting
     codes obtained from the Calderbank-Shor-Steane (CSS)
construction applied to  $(1+u)$-constacyclic  codes over
      $\mathbb{F}_{2^{m}}+u\mathbb{F}_{2^{m}}.$


\dse{1~~Introduction} Codes over finite rings have been paid much
attention from the 1990s since the landmark paper [1] showed that
certain nonlinear binary codes are images of linear codes over
$\mathbb{Z}_{4}$ under the Gray map. Constacyclic codes play a
significant role in the theory of error-correcting codes, which can
be efficiently encoded using shift registers. Constacyclic codes,
such as cyclic and negacyclic codes,  over various types of finite
rings have been studied. Constacyclic codes over finite commutative
rings were first introduced by Wolfmann in [2], where it was proved
that the binary image of a linear negacyclic code over
$\mathbb{Z}_{4}$ is a binary cyclic code (not necessarily linear).
Later, Ling and Blackford [3] extended some results in [2] to the
ring $\mathbb{Z}_{p^{k+1}}.$ Dinh and Permouth [4] gave structure
theorems for cyclic and negacyclic codes over finite chain
rings.\\

In recent years, many papers have been published dealing with codes
over polynomial residue rings. Codes over polynomial residue rings
were first introduced by Bachoc in [5].  They are shown to be
connected to lattices and have generated interest among coding
theorists since then. Cyclic codes and self-dual codes over
$\mathbb{F}_{2}+u\mathbb{F}_{2}$ were given by Bonnecaze and Udaya
[6]. Some results about cyclic codes over
$\mathbb{F}_{2}+v\mathbb{F}_{2}$ were given by Zhu et al. [7]. Type
II codes over $\mathbb{F}_{2}+u\mathbb{F}_{2}$ were given by
Dougherty et al.[8], then type II codes over
$\mathbb{F}_{2^{m}}+u\mathbb{F}_{2^{m}}$ were studied by Betsumiya
et al. [9]. Constacyclic codes over polynomial residue rings have
been received much attention. Qian et al. [10]
 generalized the main results of [2] to the ring
$\mathbb{F}_{2}+u\mathbb{F}_{2}.$  Amarra and Nemenzo [11]
generalized the results of [10] to the setting of $(1-u)$-cyclic
codes over $\mathbb{F}_{p^{k}}+u\mathbb{F}_{p^{k}}.$  Abualrub and
Siap [12] studied $(1-u)$-constacyclic codes of arbitrary length
over $\mathbb{F}_{2}+u\mathbb{F}_{2}.$ Dinh [13] gave the structure of
constacyclic codes of length $2^{s}$ over Galois extension rings of
$\mathbb{F}_{2}+u\mathbb{F}_{2}.$  Karadenniz and Yildiz [14]
studied $(1 + v)-$constacyclic codes over
$\mathbb{F}_{2}+u\mathbb{F}_{2}+v\mathbb{F}_{2}+uv\mathbb{F}_{2}.$
 Kai et al.[15] studied $(1 + u)-$ constacyclic codes over
 $\mathbb{F}_{2}+u\mathbb{F}_{2}+v\mathbb{F}_{2}+uv\mathbb{F}_{2}.$\\

In this paper, we focus on codes over the ring
$\mathbb{F}_{2^{m}}+u\mathbb{F}_{2^{m}}$, where $u^{2}=0$. As we know,
 Amarra and Nemenzo [11] have already studied  $(1-u)$-cyclic codes over
$\mathbb{F}_{p^{k}}+u\mathbb{F}_{p^{k}}.$  The ring
$\mathbb{F}_{2^{m}}+u\mathbb{F}_{2^{m}}$ is just a special ring of
$\mathbb{F}_{p^{k}}+u\mathbb{F}_{p^{k}}.$ However, in the paper of
Amarra and Nemenzo, the authors discussed  $(1-u)$-cyclic codes by
the homogeneous weight over
$\mathbb{F}_{p^{k}}+u\mathbb{F}_{p^{k}}.$ In this paper, we
investigate $(1+u)$-constacyclic codes over
$\mathbb{F}_{2^{m}}+u\mathbb{F}_{2^{m}}$ by the Lee weight.
Therefore, the Gray map in this paper is different from theirs, so
our results obtained in this paper is also not a special case of
theirs. We define a Gray map from
$\mathbb{F}_{2^{m}}+u\mathbb{F}_{2^{m}}$ to $\mathbb{F}_{2}^{2m}$
and prove that the image of a $(1+u)$-constacyclic code of odd
length $n$ over $\mathbb{F}_{2^{m}}+u\mathbb{F}_{2^{m}}$ under the
Gray map is a distance-invariant quasicyclic code of  index $m$ and
length $2mn$ over $\mathbb{F}_{2}.$ We also prove that every code of
length $2mn$ which is the Gray image of  cyclic codes over
$\mathbb{F}_{2^{m}}+u\mathbb{F}_{2^{m}}$ of length $n$ is
permutation equivalent to a binary quasi-cyclic code of index $m.$
Furthermore, we construct a family of quantum error-correcting codes
by taking advantage of
 $(1+u)$-constacyclic  codes of odd length  over
 $\mathbb{F}_{2^{m}}+u\mathbb{F}_{2^{m}}.$\\

\dse{2~~Preliminaries}
 Let $\mathbb{F}_{2^{m}}$ be the finite field with
$2^{m}$ elements. Let $R$ be the commutative ring
$\mathbb{F}_{2^{m}}+u\mathbb{F}_{2^{m}}=\{a+ub\mid a,b\in
\mathbb{F}_{2^{m}}\}$ with $u^2=0$. Then $R$ is a local ring with a
maximal ideal given by $\{0,u \};$  the set of inversible elements
 is $\{a+ub\mid a \neq 0\};$  the residue field is
$\mathbb{F}_{2^{m}}$. Addition in $R$ is given by
$(a+bu)+(c+du)=(a+c)+(b+d)u$  and multiplication is given by
$(a+bu)(c+du)=ac+(ad+bc)u$ where $a,b,c,d \in \mathbb{F}_{2^{m}}.$ A
code of length $n$ over $R$ is a nonempty subset of $R^n$, and a
code is linear over $R$ if it is an $R$-submodule of $R^n$. Given
$n$-tuple $x=(x_{1},x_{2},\cdots,x_{n}), \
y=(y_{1},y_{2},\cdots,y_{n})\in R^{n},$ the {\it{dual of the code
$C$}}, denoted by $C^\perp$, is defined by $C^\perp= \{y \in R^n \ |
\ \sum x_i y_i=0 {\rm \ for \ all \ } x \in C \}.$ A code $C$ of
length $n$ over $R$ is called dual-containing if $C^\bot \subseteq
C$, and it is called self-dual if $C=C^\bot.$ Two codes are
 permutation equivalent if one can be obtained from the other by permuting the
coordinates. Any code over $R$ is permutation equivalent to a code
$C$ with generator matrix of the form:
$$G=
\left
(\begin {array} {ccc} I_{k_{1}} & A & B \\
0 & uI_{k_{2}} & uD
\end {array}\right),$$
where  $B$ is  a matrix   over $R$ and $A, \ D$ are
$\mathbb{F}_{2^{m}}$ matrices. Then $C$ is an abelian group of type
$(4^{m})^{k_{1}}(2^{m})^{k_{2}}$, $C$ contains $2^{2mk_{1}+mk_{2}} $
codewords, and $C$ is a free $R$-module if and only if $k_{2}=0$.\\

Let $C$ be a code of length $n$ over $R$ and $P(C)$ be its
polynomial representation, i.e.,
$$P(C)=\left\{\sum_{i=0}^{n-1}c_ix^i\mid(c_0,c_1,\ldots,c_{n-1})\in C\right\}.$$
Let $\sigma$ and $\nu$ be maps from $R^n$ to $R^n$ given by
$$\sigma(c_0,c_1,\ldots,c_{n-1})=(c_{n-1}, c_0,\ldots,c_{n-2}),$$
 and
$$\nu(c_0,c_1, \ldots,c_{n-1})=((1+u)
c_{n-1},c_0,\ldots,c_{n-2}),$$ respectively. Then $C$ is said to be
cyclic if $\sigma(C)=C$ and $(1+u)$-constacyclic if $\nu(C)=C$. A
code $C$ of length $n$ over $R$ is cyclic if and only if $P(C)$ is
an ideal of $R[x] / {\langle x^n-1\rangle}$ and a code $C$ of length
$n$ over $R$ is $(1+u)$-constacyclic if and only if $P(C)$ is an
ideal of $R[x]/ {\langle x^n-(1+u)\rangle}$. Throughout this paper,
we assume that the length $n$ of $(1+u)$-constacyclic  codes  is odd.\\

\dse{3~~ Lee weights of linear codes and  Gray map over
$\mathbb{F}_{2^{m}}+u\mathbb{F}_{2^{m}}$}
 Linear  codes over the
ring $R$ were studied [9, 16]. In that work, the authors used the
notion of a trace orthogonal basis to define the Lee weight over
this ring. Now, we first give the definition of Trace Orthogonal
Basis, then we give the definition of the Lee weight for the
elements of
$R.$\\

\noindent\textbf{Definition 3.1$^{[9,16]}$.}  \emph{Let
$B=\{\alpha_{1},\alpha_{2},\cdots,\alpha_{m}\}$ be a basis for
$\mathbb{F}_{2^{m}}$ over $\mathbb{F}_{2}.$   We call that $B$ is a
trace orthogonal basis (TOB), if we have
\begin{eqnarray*}
\emph{Tr} (\alpha_{i}\cdot \alpha_{j}) = \left\{ {{\begin{array}{ll}
 {0}, & {\textrm{if}\mbox{ } \alpha_{i}\neq \alpha_{j}},\\
 {1}, & {\textrm{if}\mbox{ } \alpha_{i}= \alpha_{j}},\\
\end{array} }} \right .
\end{eqnarray*}
where $\emph{Tr}$ is the usual Trace function from
$\mathbb{F}_{2^{m}}$ to $\mathbb{F}_{2}.$}\\

\noindent\textbf{Definition 3.2$^{[9,16]}$.}\emph{ Let
$B=\{\alpha_{1},\alpha_{2},\cdots,\alpha_{m}\}$ be  a TOB of
$\mathbb{F}_{2^{m}}$ over $\mathbb{F}_{2}.$  Then for
$x=x_{1}\alpha_{1}+x_{2}\alpha_{2}+\cdots+x_{m}\alpha_{m}\in
\mathbb{F}_{2^{m}},$ the Lee weight of $x$ with respect to the basis
$B,$ denoted by $w_{L}(x),$ is defined to be the number of
$x_{i}^{,}$s that
 are non-zero. The Lee weight $w_{L}$ of $a+ub \in R$ is defined
to be the sum of the Lee weight of
$b$ and that of $a+b,$ i.e., $w_{L}(a+ub)=w_{L}(b,\ a+b).$}\\

 Define the Lee weight of a codeword
$\textbf{c}=(c_{0},c_{1},\ldots,c_{n-1})\in R^n$ to be the rational
sum of the Lee weights of its components, i.e.,
$w_{L}(\textbf{c})=\sum_{i=0}^{n-1}w_{L}(c_{i})$. For any
$c_{1},c_{2}\in R^n$, the Lee distance $d_{L}$ is given by
$d_{L}(\textbf{c}_{1},\textbf{c}_{2})=w_{L}(\textbf{c}_{1}-\textbf{c}_{2})$.
The minimum Lee distance of $C$ is the smallest nonzero Lee distance
between all pairs of distinct codewords of $C$. The minimum Lee
weight of $C$ is the smallest nonzero Lee weight among all codewords
of $C$. If $C$ is linear, then the minimum Lee distance is the same
as the minimum Lee weight. The Hamming weight $w(\textbf{c})$ of a
codeword $\textbf{c}$ is the number of nonzero components in
$\textbf{c}$. The Hamming distance
$d(\textbf{c}_{1},\textbf{c}_{2})$ between two codewords $\textbf{c}_{1}$ and
$\textbf{c}_{2}$ is the Hamming weight of the codeword
$\textbf{c}_{1}-\textbf{c}_{2}$. The minimum Hamming distance $d$ of
$C$ is
defined as $\min\{d(\textbf{c}_{1},\textbf{c}_{2})|\textbf{c}_{1},\textbf{c}_{2}\in C,\textbf{c}_{1}\neq \textbf{c}_{2}\}$.\\

Now we give the definition of the Gray map on $R^n$.  Observe that
any element $c\in R$ can be expressed as $c=r+uq$, where
$r=r_{0}\alpha_{1}+r_{1}\alpha_{2}+\cdots+r_{m-1}\alpha_{m}, \
q=q_{0}\alpha_{1}+q_{1}\alpha_{2}+\cdots+q_{m-1}\alpha_{m}\in
\mathbb{F}_{2^{m}}.$ The Gray map $\Phi:R\rightarrow
\mathbb{F}_{2}^{2m}$ is given by
$\Phi(c)=(q_{0},q_{1},\ldots,q_{m-1},r_{0}+q_{0},r_{1}+q_{1},\ldots,r_{m-1}+q_{m-1})$.
This map can be extended to $R^n$ in a natural way:
\begin{eqnarray*}
&&\Phi:R^n\rightarrow \mathbb{F}_2^{2mn}\\
&&(c_{0},c_{1},\ldots,c_{n-1})\mapsto(q_{0,1},\ldots,q_{0,m},\ldots,q_{n-1,1},\ldots,q_{n-1,m},\\
&&r_{0,1}+q_{0,1},\ldots,r_{0,m}+q_{0,m},\ldots,
r_{n-1,1}+q_{n-1,1},\ldots,r_{n-1,m}+q_{n-1,m})
\end{eqnarray*}
where $c_{i}=r_{i}+uq_{i}$ with
$r_{i}=r_{i,1}\alpha_{1}+r_{i,2}\alpha_{2}+\cdots+r_{i,m}\alpha_{m},$
\
$q_{i}=q_{i,1}\alpha_{1}+q_{i,2}\alpha_{2}+\cdots+q_{i,m}\alpha_{m}$
for
$0\le i\le n-1$.\\

It is clear that $\Phi $ preserves linearity of codes.  The
following property of the Gray map is obvious from the definitions.\\

\noindent\textbf{Proposition 3.3.} \emph{The Gray map $\Phi$ is a
distance-preserving map
from ($R^n$, Lee distance) to ($\mathbb{F}_{2}^{2mn}$, Hamming distance) and it is also $\mathbb{F}_2$-linear.}\\

\dse{4~~Gray images of  $(1+u)$-constacyclic codes over
$\mathbb{F}_{2^{m}}+u\mathbb{F}_{2^{m}}$}

Let $\sigma^{m}$ be the map from $\mathbb{F}_{2}^{2mn}$ to
$\mathbb{F}_{2}^{2mn}$ given by $$\sigma^{m}(a)=(a^{(2n-1)} |
a^{(0)} | \cdots |a^{(2n-2)}), $$  where $a^{(i)} \in
(\mathbb{F}_{2})^{m},$ for all $ i=0,1,\ldots ,2n-1,$ and $ |$
denotes the usual vector concatenation. Let $C$ be a code of length
$2mn$ over $\mathbb{F}_{2},$ if $\sigma^{ m}(C)=C, $ then the code
$C$ is said to be a quasi-cyclic code of index
$m.$\\

\noindent\textbf{Proposition 4.1.} \emph{Let $\nu$ denote the
$(1+u)$-constacyclic shift of $R^n$ and $\sigma$ the cyclic shift of
$\mathbb{F}_{2}^{2mn}$. Let $\Phi$ be the Gray map of $R^n$ into
$\mathbb{F}_{2}^{2mn}$.
Then $\Phi\nu=\sigma^{ m}\Phi$.}\\

\noindent\textbf{Proof.} Let
$\textbf{c}=(c_{0},c_{1},\ldots,c_{n-1})\in R^n$, where
$c_{i}=r_{i}+uq_{i}$ with
$r_{i}=r_{i,1}\alpha_{1}+r_{i,2}\alpha_{2}+\cdots+r_{i,m}\alpha_{m},$
\
$q_{i}=q_{i,1}\alpha_{1}+q_{i,2}\alpha_{2}+\cdots+q_{i,m}\alpha_{m}$
for $0\leq i \leq n-1.$  From the definition of the Gray map, we
obtain
\begin{eqnarray*}
\Phi(\textbf{c})&=&(q_{0,1},\ldots,q_{0,m},\ldots,q_{n-1,1},\ldots,q_{n-1,m},
r_{0,1}+q_{0,1}, \ldots,\\
&&r_{0,m}+q_{0,m},\ldots,
r_{n-1,1}+q_{n-1,1},\ldots,r_{n-1,m}+q_{n-1,m}).
\end{eqnarray*}
Hence
\begin{eqnarray*}
\sigma^{
m}(\Phi(\textbf{c}))&=&(r_{n-1,1}+q_{n-1,1},\ldots,r_{n-1,m}+q_{n-1,m},q_{0,1},\ldots,q_{0,m},\ldots,q_{n-1,1},\ldots,\\
&&q_{n-1,m}, r_{0,1}+q_{0,1},\ldots, r_{0,m}+q_{0,m},\ldots,
r_{n-2,1}+q_{n-2,1},\ldots,r_{n-2,m}+q_{n-2,m}).
\end{eqnarray*}
On the other hand
\begin{eqnarray*}
\nu(\textbf{c})& = &((1+u)c_{n-1},c_{0},\ldots,c_{n-2})\\
       & = &(r_{n-1}+u(r_{n-1}+q_{n-1}),r_{0}+uq_{0},\ldots,r_{n-2}+uq_{n-2}).
\end{eqnarray*}
We can deduce that
\begin{eqnarray*}
\Phi(\nu(\textbf{c}))&=&(r_{n-1,1}+q_{n-1,1},\ldots,r_{n-1,m}+q_{n-1,m},q_{0,1},\ldots,q_{0,m},\ldots,q_{n-1,1},\ldots,\\
&&q_{n-1,m}, r_{0,1}+q_{0,1}, \ldots,r_{0,m}+q_{0,m},\ldots,
r_{n-2,1}+q_{n-2,1},\ldots,r_{n-2,m}+q_{n-2,m}).
\end{eqnarray*}
Therefore   $$\Phi\nu=\sigma^{ m}\Phi.    \ \ \ \ \ \ \ \ \ \ \ \ \qed$$\\

\noindent\textbf{Theorem 4.2.} \emph{A linear code $C$ of length $n$
over $R$ is a $(1+u)$-constacyclic code if and only if $\Phi(C)$ is a  quasi-cyclic code of index $m$ and length $2mn$ over $\mathbb{F}_{2}$.}\\

{\it Proof}~~  It is an immediate consequence of Proposition 4.1.\qed\\

\noindent Consequently, we can obtain the following corollary.\\

\noindent\textbf{Corollary 4.3.} \emph{The Gray image of a
$(1+u)$-constacyclic code of length $n$ over $R$ under the Gray map
$\Phi$ is a distance-invariant a  linear quasi-cyclic code of index
$m$ and length $2mn$ over $\mathbb{F}_{2}$.}\\

{\it Proof}~~ Suppose that $C$ is a $(1+u)$-constacyclic code.
Then, by Proposition 4.1, we get
$$\sigma^{
m} \Phi(C)=\Phi\nu(C)=\Phi(C).$$ Hence, $\Phi(C)$ is a quasi-cyclic
code of index
$m.$ \\
Conversely,  if $\Phi(C)$ is a quasi-cyclic code of index $m,$ then,
by Proposition 4.1, we obtain
$$\Phi\nu(C)=\sigma^{
m}\Phi(C)=\Phi(C).$$
 It follows that $\nu(C)=C.$ $\Box$\\

\dse{5~~Gray images of cyclic codes over
$\mathbb{F}_{2^{m}}+u\mathbb{F}_{2^{m}}$ of odd length } {\upshape
Note that $(1 + u)^{n} = 1 + u$ if $n$ is odd and $(1 + u)^{n} = 1$
if $n$ is even. We will study the properties of $(1 +
u)$-constacyclic codes of odd
length in this section. First, we give some definitions.}\\

Define the map $\mu$ as
\begin{eqnarray*}
&&\mu: \ R [x]/ (x^{n}-1)\rightarrow R [x]/
(x^{n}-(1+u)),\\
&& \ \ \ \ \ \ c(x)\mapsto c((1+u) x).
\end{eqnarray*}
We know that if  $n$ is odd, then $\mu$ is a ring isomorphism. Hence
$I$ is an ideal of $R [x]/ (x^{n}-1)$ if and only if $\mu(I)$ is an
ideal
of $R [x]/ (x^{n}-(1+u)).$\\
Consequently, $\bar{\mu}$ is the map
\begin{eqnarray*}
&&\bar{\mu}: \ R^{n} \rightarrow
R^{n},\\
&&(c_{0},c_{1},\ldots,c_{n-1})\mapsto (c_{0},(1+u)
c_{1},\ldots,(1+u)^{n-1}c_{n-1}).
\end{eqnarray*}

Hence, we get the following theorem.\\
 \noindent\textbf{\upshape
Theorem 5.1.\label{ortho}}~~\emph{The set $C\subseteq R^{n}$ is a
cyclic code if and only if $\bar{\mu}(C)$ is a
$(1 +u)$-constacyclic code.}\\

Let $\textbf{c}=(c_{0},c_{1},\ldots,c_{n-1})\in R^n$, where
$c_{i}=r_{i}+uq_{i}$ with
$r_{i}=r_{i,1}\alpha_{1}+r_{i,2}\alpha_{2}+\cdots+r_{i,m}\alpha_{m},$
\
$q_{i}=q_{i,1}\alpha_{1}+q_{i,2}\alpha_{2}+\cdots+q_{i,m}\alpha_{m}$
for $0\leq i \leq n-1,$  and
\begin{eqnarray*}
&&\Phi(\textbf{c})=(q_{0,1},\ldots,q_{0,m},\ldots,q_{n-1,1},\ldots,q_{n-1,m},
r_{0,1}+q_{0,1}, \\
&&r_{0,m}+q_{0,m},\ldots,
r_{n-1,1}+q_{n-1,1},\ldots,r_{n-1,m}+q_{n-1,m}).
\end{eqnarray*}
In the following, we simply denote \begin{eqnarray*}
\Phi(\textbf{c})=(c_{\langle 0 \rangle},c_{\langle 1
\rangle},\ldots,c_{\langle n-1 \rangle},c_{\langle n
\rangle},c_{\langle n+1 \rangle},\ldots,c_{\langle 2n-1 \rangle}),
\end{eqnarray*}
where $c_{\langle i \rangle}=(q_{i,1},q_{i,2},\ldots, q_{i,m});
c_{\langle n+i \rangle}=(q_{i,1}+r_{i,1},q_{i,2}+r_{i,2},\ldots,
q_{i,m}+r_{i,m}),$ for $0 \leq i \leq n-1 .$

\noindent\textbf{\upshape Definition 5.2.\label{ortho}}~~Let $\tau$
be the following permutation of $\{\langle 0 \rangle, \langle 1
\rangle ,\ldots, \langle 2n-1 \rangle\}$ with $n$ odd:
$$\tau=(\langle 1 \rangle, \langle n+1 \rangle)(\langle 3 \rangle, \langle n+3 \rangle) \cdots(\langle 2i+1 \rangle, \langle n+2i+1 \rangle)
\cdots (\langle n-2 \rangle, \langle 2n-2 \rangle).$$ The Nechaev
permutation is the permutation $\pi$ of
$\mathbb{F}_{2}^{2mn}$ defined as\\
$$\pi^{\otimes m}(\textbf{c})=(c_{\tau(\langle 0 \rangle)},c_{\tau(\langle 1 \rangle)},\ldots,
c_{\tau(\langle 2n-1 \rangle)}).$$

\noindent\textbf{\upshape Theorem
5.3.\label{ortho}}~~\emph{$\Phi\bar{\mu}=\pi^{\otimes m} \Phi.$}

{\it Proof}~~ Let $\textbf{c}=(c_{0},c_{1},\ldots,c_{n-1})\in R^n$,
where $c_{i}=r_{i}+uq_{i}$ with
$r_{i}=r_{i,1}\alpha_{1}+r_{i,2}\alpha_{2}+\cdots+r_{i,m}\alpha_{m},$
\
$q_{i}=q_{i,1}\alpha_{1}+q_{i,2}\alpha_{2}+\cdots+q_{i,m}\alpha_{m}$
for $0\leq i \leq n-1.$  Then
\begin{eqnarray*}
\Phi(\textbf{c})=(c_{\langle 0 \rangle},c_{\langle 1
\rangle},\ldots,c_{\langle n-1 \rangle},c_{\langle n
\rangle},c_{\langle n+1 \rangle},\ldots,c_{\langle 2n-1 \rangle}).
\end{eqnarray*}
On the other hand
$$\bar{\mu}(\textbf{c})=(c_{0},(1+u) c_{1},\ldots,(1+u)^{n-1}c_{n-1}),$$
$$\Phi(\bar{\mu}(\textbf{c}))=(s_{\langle 0 \rangle},s_{\langle 1 \rangle},\ldots,s_{\langle n-1 \rangle},s_{\langle n \rangle},s_{\langle n+1\rangle},
\ldots,s_{\langle 2n-1 \rangle}),$$
then for $0\leq j\leq n-1$\\
if $j$ is even, then $s_{\langle j \rangle}=(q_{j,1},q_{j,2},\ldots,
q_{j,m}),$ $s_{\langle n+j
\rangle}=(q_{j,1}+r_{j,1},q_{j,2}+r_{j,2},\ldots,
q_{j,m}+r_{j,m});$\\
if $j$ is odd, then $s_{\langle j
\rangle}=(q_{j,1}+r_{j,1},q_{j,2}+r_{j,2},\ldots,
q_{j,m}+r_{j,m}),$ $s_{\langle n+j
\rangle}=(q_{j,1},q_{j,2},\ldots, q_{j,m}).$ Therefore
 $$\pi^{\otimes m}
\Phi(\textbf{c})=\Phi \bar{\mu}(\textbf{c}). \ \ \  \ \ \ \ \ \Box$$

From Theorem 5.3, the following corollary can be obtained.\\
\textbf{\upshape Corollary 5.4.\label{ortho}}~~\emph{Let $\tilde{C}$
is the Gray image of  cyclic codes of length $n$ over $R.$ Then
$\tilde{C}$ is equivalent to a quasi-cyclic code of index $m$ and
length $2mn$ over $\mathbb{F}_{2}.$}

{\it Proof}~~From Theorem 5.1, a code  $C$ is a cyclic code over $
R$ if and only if $\bar{\mu}(C)$ is a $(1+u)$-constacyclic codes.
From Theorem 4.2, the code is cyclic if and only if its Gray image
is a quasi-cyclic code of index $m$ over $\mathbb{F}_{2},$ that is,
$\pi^{\otimes m} \Phi(C)$ is a quasi-cyclic code of index $m$ over
$\mathbb{F}_{2}.$  $\Box$ \\

Now, we give an example to illustrate the above results as
follows. Comparing with the known binary linear  code in Ref. 17, our obtained binary linear codes are optimal.   \\
 \noindent\textbf{Example 5.5.} Let
 $\mathbb{F}_{4}=\{0,1,\omega,\omega^{2}=1+\omega\}.$ Consider $(1+u)$-constacyclic codes over $\mathbb{F}_{4}+u\mathbb{F}_{4}$
of length $n=3.$  Note

 \begin{eqnarray*}
 x^{3}-1=(x +1)(x+\omega )(x+\omega^{2}).
 \end{eqnarray*}
 Applying the map $\mu : \ x\mapsto (1+u)x,$ we obtain
 \begin{eqnarray*}
 x^{3}-(1+u)=(x +1+u)(x+\omega
 (1+u))(x+\omega^{2}(1+u))=f(x)g(x)h(x).
 \end{eqnarray*}
 Let $C_{1}$ be the $(1+u)$-constacyclic codes over
 $\mathbb{F}_{4}+u\mathbb{F}_{4}$ of length 3 with generator
 polynomial
 $$C_{1}=\langle uf(x)g(x)\rangle .$$ The Gray image $\Phi(C_{1})$ is a
 $[12,2,8]$- binary linear codes, which is an optimal code.\\
 Let $C_{2}$ be the $(1+u)$-constacyclic codes over
 $\mathbb{F}_{4}+u\mathbb{F}_{4}$ of length 3 with generator
 polynomial
 $$C_{2}=\langle f(x)g(x), u f(x)h(x) \rangle.$$
 The Gray image $\Phi(C_{2})$ is a $[12,6,4]$-binary
 linear code, which is an optimal code.\\

Now, we consider $(1+u)$-constacyclic codes over
$\mathbb{F}_{4}+u\mathbb{F}_{4}$ of length $n=5.$  Note

 \begin{eqnarray*}
 x^{5}-1=(x +1)(x^2+\omega x+1 )(x^2+\omega^{2}x+1).
 \end{eqnarray*}
 Applying the map $\mu : \ x\mapsto (1+u)x,$ we obtain
 \begin{eqnarray*}
 x^{5}-(1+u)=(x +1+u)((1+u)x^2+\omega x+1+u)((1+u)x^2+\omega^{2}x+1+u)=f(x)g(x)h(x).
 \end{eqnarray*}
 Let $C$ be the $(1+u)$-constacyclic codes over
 $\mathbb{F}_{4}+u\mathbb{F}_{4}$ of length 5 with generator
 polynomial
 $$C=\langle f(x)g(x), u f(x)h(x) \rangle.$$
 The Gray image $\Phi(C)$ is a $[20, 12, 4]$-binary
 linear code, which is an optimal code.\\

\dse{6~~An Application to $(1+u)$-constacyclic  codes over the ring $R$}

Quantum error-correcting codes play an important role not only in
quantum communication but also quantum computation[18]. Many good
quantum error-correcting codes have been constructed by using codes
over finite rings([19-24]).    As an application of the previous
results,  we will construct a family of quantum error-correcting
codes by taking advantage of the Calderbank-Shor-Steane (CSS)
construction applied to $(1+u)$-constacyclic  codes of odd length
$n$ over $R=\mathbb{F}_{2^{m}}+u\mathbb{F}_{2^{m}}.$ We first give
the following
definitions and notations.\\

 Let $f(x)=a_{k}x^{k}+a_{k-1}x^{k-1}+\cdots+a_{0}$ be a
polynomial in $R[x],$  where $a_{0}$ is an invertible element in
$R.$ Define the reciprocal polynomial of $f(x)$ as
$f^{*}(x)=a_{0}^{-1}x^{k}f(x^{-1})$, i.e.,
$f^{*}(x)=x^{k}+b_{1}x^{k-1}+\cdots+b_{k},$ where
$b_{i}=a_{0}^{-1}a_{i},$ for all $i=1, 2, \ldots, k.$ Obviously,
$(f^{*}(x))^{*}=f(x)$ and $(f(x)g(x))^{*}=f^{*}(x)g^{*}(x)$.  Let
$\mathcal{C}$ be a cyclic code of length $n$  over $ R.$ Then, there
are unique monic polynomials $f(x), g(x)$ and $h(x)$ such that
$\mathcal{C} =\left\langle f(x)h(x),uf(x)g(x)\right\rangle$, where
$f(x)g(x) h(x)=x^{n}-1$ and $|\mathcal{C}|=
2^{m(2deg(g(x))+deg(h(x)))}$. Moreover,
$\mathcal{C^{\bot}}=\left\langle g^{*}(x)h^{*}(x),
ug^{*}(x)f^{*}(x)\right\rangle$ and $|\mathcal{C^{\bot}}|=
2^{m(2deg(f(x))+deg(h(x)))}$ [4].  It is known that the map $\mu$ is
a ring isomorphism in Section 5.  Therefore,  every
$(1+u)$-constacyclic  code $C$ has similar results as
in the Ref. [4].\\

\noindent\textbf{Theorem 6.1.} \emph{Let $C$ be a
$(1+u)$-constacyclic code over $R$ of length $n$. Then there exists
a unique family of monic pairwise coprime polynomials $f(x), g(x),
h(x)$ in $R[x]$ such that $f(x)g(x) h(x)=x^{n}-(1+u)$ and
$$C=\left\langle
f(x)h(x),uf(x)g(x)\right\rangle$$ with $|C|=
2^{m(2deg(g(x))+deg(h(x)))}$. Moreover,
$$C^{\bot}=\left\langle g^{*}(x)h^{*}(x),
ug^{*}(x)f^{*}(x)\right\rangle$$ with
$|C^{\bot}|= 2^{m(2deg(f(x))+deg(h(x)))}$. }\\

Now we obtain a sufficient and necessary condition for the existence
of dual-containing cyclic codes over $R$ by using  generator
polynomials of $(1+u)$-constacyclic codes over $R.$ The following Theorem is improved the main result (Theorem 4.2) in Ref.23. \\

 \noindent\textbf{Theorem
6.2.} \emph{ Let $C$ be a $(1+u)$-constacyclic code over $R$ of length $n$. Then
there exists a unique family of monic pairwise coprime polynomials
$f(x),g(x),h(x)$ in $R[x]$ such that $f(x)g(x)h(x)=x^{n}-(1+u)$ and
$$C=\left\langle
f(x)h(x),uf(x)g(x)\right\rangle.$$
 Then  $ C^{\bot} \subseteq C $ if and only
 if $f(x)$ divides $g^{*}(x).$
 }\\

\noindent\pf  By Theorem 6.1, we have
$$C^{\bot}=\left\langle g^{*}(x)h^{*}(x),
ug^{*}(x)f^{*}(x)\right\rangle.$$ If $ C^{\bot} \subseteq C ,$ then
there exists $a(x)\in\mathbb{F}_{2^{m}}[x]$ such that
$g^{*}(x)h^{*}(x)=f(x)h(x)a(x).$ Then $ g^{*}(x)h^{*}(x)g(x)$
 $=f(x)g(x)h(x)a(x)=$  $
-f^{*}(x)g^{*}(x)h^{*}(x)a(x),$  Therefore $f^{*}(x)$ divides
$g(x);$ that is, $f(x)$ divides $g^{*}(x).$

 To prove the necessity,  we suppose that $r(x)$ is the product of non-self-reciprocal
irreducible polynomials in $h(x)$ which do not occur in pairs. Then,
$h(x)$ can be expressed in the form $h(x) = b(x)r(x),$ where $b(x) =
b^{*}(x).$ Since $f(x)$ divides $g^{*}(x),$ then there exists $l(x)
\in \mathbb{F}_{2^{m}}[x]$ such that $g^{*}(x)=f(x)l(x).$ Obviously,
$r(x)|f^{*}(x)g^{*}(x)h^{*}(x).$ Suppose that
$\textrm{gcd}(f^{*}(x), r(x))=m(x).$  Then $m(x)$ divides
$f^{*}(x)$, which means $m(x)$ divides $g(x).$  On the other hand,
$m(x)$ divides $r(x)$, so $m(x)$ divides $h(x).$ Since $g(x)$ and
 $h(x)$ are relatively coprime, it follows that $m(x)=1.$  This shows that $r(x)$ and $f^{*}(x)$ are relatively
prime. Also, $r(x)$ and $h^{*}(x)$ are relatively prime. It follows
that $r(x)$ divides $g^{*}(x)$. Hence, $f(x)r(x)$ divides
$g^{*}(x)$. Let $g^{*}(x) = f(x)r(x)k(x),$ for some $k(x)
\in\mathbb{F}_{2^{m}}[x].$ Since $g(x)$ and
 $h(x)$ are relatively coprime, then there exist $s(x)$ and $t(x)$ in
 $\mathbb{F}_{2^{m}}[x]$ such that $g(x)s(x)+h(x)t(x)=1.$ Therefore
\begin{eqnarray*}
 uf^{*}(x)g^{*}(x)&=& uf^{*}(x)g^{*}(x)(g(x)s(x)+h(x)t(x))\\
&=& uf^{*}(x)f(x)r(x)k(x)g(x)s(x) +uf^{*}(x)f(x)r(x)k(x)h(x)t(x) \in
C.
\end{eqnarray*}
 Furthermore
\begin{eqnarray*}
g^{*}(x)h^{*}(x)
&=&f(x)r(x)m(x)b^{*}(x)r^{*}(x)\\
&=&  f(x)h(x)k(x)r^{*}(x)\in C.
\end{eqnarray*}
Hence $ C^{\bot} \subseteq C ,$ which
completes the proof.\qed\\

\noindent\textbf{Remark. }  Theorem 6.2 can be generalized directly
to cyclic codes over finite chain rings.\\

 A fundamental link between
linear codes and quantum codes is given by the
Calderbank-Shor-Steane(CSS) construction.\\
\noindent\textbf{Theorem  6.3$^{[18]}$. (CSS Construction)}
\emph{Let $C$ and $C'$ be two binary codes with parameters $[n,k_{1}
,d_{1} ]$ and $[n,k_{2} ,d_{2} ],$ respectively. If
$C^{\perp}\subseteq C',$ then an $[[n,k_{1}+ k_{2}-n, \emph{min}
\{d_{1},d_{2}\}]]_{2}$ quantum code can be constructed. Especially,
if $C^{\perp}\subseteq C,$ then there exists an
$[[n,2k_{1}-n,d_{1}]]_{2}$
quantum code.}\\

Now, based on dual-containing $(1+u)$-constacyclic  codes over $R,$
we construct a class of binary quantum code by using the
Calderbank-Shor-Steane(CSS) construction.  We must need  the
following theorems for analyzing
 construction a new family binary quantum codes from dual-containing $(1+u)$-constacyclic
 codes over
 $R.$\\
\noindent\textbf{Theorem 6.4.} \emph{Let $C$ be  a dual-containing
$(1+u)$-constacyclic  code of length $n$  over  $R$.
Then $\Phi(C)$ is a dual-containing linear quasi-cyclic code of length $2mn$ over $\mathbb{F}_{2}$.}\\
{\it Proof}~~Suppose that  any $ c_{1}=r_{1}+uq_{1}\in C^{\bot},\
c_{2}=r_{2}+uq_{2}\in C$
 with
 $r_{i}=r_{i,1}\alpha_{1}+r_{i,2}\alpha_{2}+\cdots+r_{i,m}\alpha_{m},$
 $q_{i}=q_{i,1}\alpha_{1}+q_{i,2}\alpha_{2}+\cdots+q_{i,m}\alpha_{m},$ for $1 \leq i \leq 2.$  Since $C$  contains its dual,
then $c_{1}\cdot c_{2}=0$; that is, $r_{1} r_{2}=0$ and $r_{1}
q_{1}+ r_{2}q_{2}=0.$ Hence, $r_{1}
r_{2}=\sum_{i,j}(r_{1,i}r_{2,j})\alpha_{i}\alpha_{j}=0$ and $r_{1}
q_{1}+
r_{2}q_{2}=\sum_{i,j}(r_{1,i}q_{1,j}+r_{2,i}q_{2,j})\alpha_{i}\alpha_{j}=0.$
Taking the trace over $\mathbb{F}_{2},$   we get
$$\sum_{i,j}(r_{1,i}r_{2,j})\textrm{Tr}(\alpha_{i}\alpha_{j})=0, \ \
\sum_{i,j}(r_{1,i}q_{1,j}+r_{2,i}q_{2,j})\textrm{Tr}(\alpha_{i}\alpha_{j})=0.$$
Since $B=\{\alpha_{1},\alpha_{2},\cdots,\alpha_{m}\}$ is a trace
orthogonal basis, then $\textrm{Tr}(\alpha_{i}^{2})=1,$ for  $i=j;$
$\textrm{Tr}(\alpha_{i}\alpha_{j})=0,$ for  $i\neq j.$ Therefore,
 $\sum_{i=j}(r_{1,i}r_{2,j})=0, \ \
 \sum_{i=j}(r_{1,i}q_{1,j}+r_{2,i}q_{2,j})=0.$\\
On the other hand,
 \begin{eqnarray*}
     \Phi(c_{1})\cdot \Phi(c_{2})&=&(q_{1,1},\ldots,q_{1,m},r_{1,1}+q_{1,1},\ldots,r_{1,m}+q_{1,m})
     (q_{2,1},\ldots,q_{2,m},r_{2,1}+q_{2,1},\ldots,r_{2,m}+q_{2,m})\\
&=&0.
\end{eqnarray*}
This implies that $\Phi(C^{\perp})\subseteq \Phi(C)^{\perp}.$  Now
it is enough to show that the two sets have the same cardinality.
Let $C$ be a $(1+u)$-constacyclic code with the  parameters
$[n,4^{mk_{1}}2^{mk_{2}}]$ over $R.$ Then  $C^{\bot}$ is a
$(1+u)$-constacyclic code with the  parameters
$[n,4^{m(n-k_{1}-k_{2})}2^{mk_{2}}]$ over $R.$   By Corollary 4.3,
$\Phi(C)$ is a binary $[2mn,2mk_{1}+mk_{2}]$ linear quasi-cyclic
code and $\Phi(C^{\bot})$ is a binary $[2mn,2mn-2mk_{1}-mk_{2}]$
linear quasi-cyclic code. Therefore,
$|\Phi(C)^{\perp}|=2^{2mn-2mk_{1}-mk_{2}}=|\Phi(C^{\perp})|.$ $\Box$ \\

Now, based on  dual-containing cyclic codes over $R,$ the Gray map,
 and the CSS construction, we construct a new
family
 of binary  quantum codes. From Theorems 6.2
and 6.4,  we  directly get the following results.\\

\noindent\textbf{Theorem 6.5.} \emph{Let $C=\left\langle
f(x)h(x),uf(x)g(x)\right\rangle$ be a
 $(1+u)$-constacyclic code with the  parameters $[n,4^{mk_{1}}2^{mk_{2}},d_{L}]$ over  $R,$  where $f(x)g(x)h(x)=x^{n}-(1+u).$ If
$f(x)$ divides $g^{*}(x),$ then, $C^{\perp}\subseteq C ,$  and there
exists a binary quantum error-correcting code
with parameters $[[2mn,4mk_{1}+2mk_{2}-2mn,d_{L}]]_{2}$.}\\

Next, let us use an  example to illustrate our construction method.
We are with the help of the computer algebra system MAGMA to find
good and new quantum
codes.\\
 \noindent\textbf{Example 6.6.}  Construction  two
quantum error-correcting codes from   $(1+u)$-constacyclic codes
over $\mathbb{F}_{4}+u\mathbb{F}_{4}$ of length $85$ and $93,$
respectively.  First,
\begin{eqnarray*}
 x^{85}-1&=&(x + 1)(x^2 + \omega x + 1)(x^2 + \omega^2x + 1)(x^4 + x^2 + \omega x + 1)(x^4 + \omega^2x^3 + x^2 + \omega^2x + 1)\\
 &&(x^4 + \omega x^2 + \omega^2x + 1)(x^4 + \omega^2x^2 + \omega x + 1)(x^4 + x^3 + \omega x + 1)(x^4 + x^3 + \omega^2x + 1)\\
 &&(x^4 + x^3 + \omega x^2 + x + 1)(x^4 + x^3 + \omega^2x^2 + x + 1)(x^4 + \omega x^3 + \omega^2x^2 + \omega^2x + 1)\\
 &&(x^4 + \omega x^3 + \omega^2x^2 + 1)(x^4 + \omega^2x^3 + \omega^2x^2 + \omega x + 1)(x^4 + \omega^2x^3 + x + 1)\\
 &&(x^4 + \omega x^3 + x + 1)(x^4 + \omega^2x^3 + x^2 + 1)(x^4 + \omega x^3 + \omega x^2 + \omega^2x + 1)\\
 &&(x^4 + \omega x^3 + x^2 + 1)(x^4 + \omega x^3 + x^2 + \omega x + 1)(x^4 + x^2 + \omega^2x + 1)\\
 &&(x^4 + \omega^2x^3 + \omega x^2 + 1)(x^4 + \omega^2x^3 + \omega x^2 + \omega x + 1).
 \end{eqnarray*}
Applying the map $\mu : \ x\mapsto (1+u)x$, we have
\begin{eqnarray*}
 x^{85}-(1+u)&=&(x +1+u)((1+u)x^2 + \omega x + 1+u)((1+u)x^4 + (1+u)x^2 + \omega x + 1+u)\\
 &&((1+u)x^4 + \omega x^3 + (1+u)\omega^2x^2 + \omega^2x + 1+u)((1+u)x^4 + \omega x^3 + (1+u)\omega^2x^2 + 1)\\
 &&((1+u)x^4 + \omega^2x^3 + (1+u)\omega x^2 + 1+u)((1+u)x^4 + \omega^2x^3 + \omega(1+u)x^2 + \omega x + 1+u)\\
 &&((1+u)x^4 + \omega x^3 + \omega(1+u)x^2 + \omega^2x + 1+u)((1+u)x^4 + \omega x^3 + (1+u)x^2 + 1+u)\\
 &&((1+u)x^4 + \omega x^3 + (1+u)x^2 + \omega x + 1+u)((1+u)x^4 + (1+u)x^2 + \omega^2x + 1+u)\\
 &&((1+u)x^4 + \omega^2x^3 + (1+u)\omega^2x^2 + \omega x + 1+u)((1+u)x^4 + \omega^2x^3 + x + 1+u)\\
 &&((1+u)x^4 + (1+u)\omega x^2 + \omega^2x + 1+u)((1+u)x^4 + \omega^2(1+u)x^2 + \omega x + 1+u)\\
 &&((1+u)x^4 + x^3 + (1+u)\omega x^2 + x + 1+u)((1+u)x^4 + x^3 + \omega^2x + 1+u)\\
 &&((1+u)x^4 + x^3 + \omega x + 1+u)((1+u)x^4 + x^3 +(1+u)\omega^2x^2 + x + 1+u)\\
 &&((1+u)x^2 + \omega^2x + 1+u)((1+u)x^4 + \omega^2x^3 +(1+u)x^2 + \omega^2x + 1+u)\\
 &&((1+u)x^4 + \omega x^3 + x + 1+u)((1+u)x^4 + \omega^2x^3 + (1+u)x^2 + 1+u).
 \end{eqnarray*}
Let $f(x)=((1+u)x^4+\omega x^3+\omega
(1+u)x^2+\omega^2x+1+u)((1+u)x^4+\omega
x^3+\omega^2(1+u)x^2+\omega^2x+1+u), h(x)=1,
g(x)=x^{85}-(1+u)/f(x)h(x)$, and
$$C=\left\langle f(x)h(x), uf(x)g(x)\right\rangle.$$ Then, $C$ is a
$(1+u)$-constacyclic code  over $\mathbb{F}_{4}+u\mathbb{F}_{4}$,
with the  parameters $[85,4^{154},5]$. Since $C^{\perp}\subseteq C$,
we have a $Q=[[340, 276, 5]]_{2}$ quantum code from Theorem 6.5.
 Comparing with the known quantum code $[[340, 276, 5]]_{2}$ given in Refs. 19 and 25, our obtained quantum code is with the same parameters.\\

Second, taking $n=93$, we have
 \begin{eqnarray*}
 x^{93}-1&=&(x +1)(x + \omega)(x + \omega^2)(x^5 + x^2 + 1)(x^5 + x^2 + \omega)(x^5 + x^3 + 1)(x^5 + \omega x^3 + \omega)\\
 &&((x^5 + \omega^2x^3 + \omega^2)(x^5 + \omega^2x^4 + \omega x^3 + \omega^2x + \omega)(x^5 + \omega^2x^4 + x^2 + \omega^2x + \omega)\\
  &&(x^5 + x^4 + x^3 + x + 1)(x^5 + x^4 + x^3 + x^2 + 1)(x^5 + \omega x^3 + x^2 + \omega^2x + \omega)\\
 &&(x^5 + x^4 + x^2 + x + 1)(x^5 + x^3 + x^2 + x + 1)(x^5 + \omega^2x^4 + \omega x^3 + x^2 + \omega)\\
 &&(x^5 + \omega x^4 + x^2 + \omega x + \omega^2)(x^5 + \omega x^4 + \omega^2x^3 + x^2 + \omega^2)(x^5 + x^2 + \omega^2)\\
 &&(x^5 + \omega^2x^3 + x^2 + \omega x + w^2)(x^5 + \omega x^4 + \omega^2x^3 + \omega x + \omega^2).
 \end{eqnarray*}
 Applying the map $\mu : \ x\mapsto (1+u)x,$ we obtain
 \begin{eqnarray*}
 x^{93}-(1+u)&=&(x +1+u)(x + \omega(1+u))(x + \omega^2(1+u))(x^5 + (1+u)x^2 + 1+u)\\
 &&(x^5 + (1+u)x^2 + \omega(1+u))(x^5 + (1+u)x^2 + \omega^2(1+u))(x^5 + \omega^2x^3+\omega^2(1+u))\\
 &&(x^5 + (1+u)x^4 + x^3 + (1+u)x^2 + 1+u)(x^5 + x^3 + 1+u)(x^5 + \omega^2x^3+(1+u)x^2\\
 && + \omega x + \omega^2(1+u))(x^5 + \omega x^3 + \omega(1+u))(x^5 + (1+u)x^4 + (1+u)x^2 + x+ 1+u)\\
 &&(x^5 + x^3 + (1+u)x^2 + x + 1+u)(x^5 + (1+u)x^4 + x^3 + x + 1+u)(x^5 + \omega x^3\\
 && + (1+u)x^2 + \omega^2x + \omega(1+u))(x^5 + \omega(1+u)x^4 + (1+u)x^2 + \omega x + \omega^2(1+u))\\
 &&(x^5 + \omega(1+u)x^4 + \omega^2x^3+ \omega x + \omega^2(1+u))(x^5 + \omega(1+u)x^4 + \omega^2x^3 +(1+u)x^2\\
 && + \omega^2(1+u))(x^5 + \omega^2(1+u)x^4 + (1+u)x^2 + \omega^2x + \omega(1+u))(x^5 + \omega^2(1+u)x^4\\
 && + \omega x^3 + \omega^2x + \omega(1+u))(x^5 + \omega^2(1+u)x^4 + \omega x^3 + (1+u)x^2 +\omega(1+u)).
 \end{eqnarray*}
Let $f(x)=(x^5+\omega^2(1+u)x^4+\omega
x^3+\omega^2x+\omega(1+u))(x+\omega(1+u)),
h(x)=x^5+(1+u)x^4+x^3+(1+u)x^2 +1+u, g(x)=x^{93}-(1+u)/ f(x)h(x)$,
and $$C=\left\langle f(x)h(x), uf(x)g(x)\right\rangle.$$ Then, $C$
is a $(1+u)$-constacyclic code over
$\mathbb{F}_{4}+u\mathbb{F}_{4}$, with  the  parameters
$[93,4^{164}2^{10},5]$.
 Since $C^{\perp}\subseteq C$, we have a $Q'=[[372, 304, 5]]_{2}$ quantum code from Theorem 6.5. Comparing with the known quantum code $[[372, 297, 5]]_2$ given in Ref. 25, our obtained quantum code is  optimal and new.\\

\dse{7~~Conclusion} In this paper, we   introduce
$(1+u)$-constacyclic codes of odd length over
$\mathbb{F}_{2^{m}}+u\mathbb{F}_{2^{m}}$ and study  their Gray
images.  We find that some optimal binary linear codes  can be
derived from such codes.  Furthermore, $(1+u)$-constacyclic codes of
odd length over $\mathbb{F}_{2^{m}}+u\mathbb{F}_{2^{m}}$ are applied
to construct binary  quantum codes.  We also  find  that some
optimal and new binary quantum codes  can be obtained from such
codes.   It would be very interesting to find  more good  linear
codes and quantum codes  from other classes  of constacyclic codes
over finite rings.

\end{document}